\documentclass[aps,pra,reprint,amsmath,amssymb,floatfix,superscriptaddress]{revtex4-2}

\usepackage{graphicx}
\usepackage{dcolumn}
\usepackage{bm}

\newcommand{\be}{\begin{eqnarray}}
\newcommand{\ee}{\end{eqnarray}}
\newcommand{\ket}[1]{\ensuremath{\left| {#1} \right>}}
\newcommand{\bra}[1]{\ensuremath{\left< {#1} \right|}}

\newcommand{\affD}{Institute for Quantum Electronics, ETH Zürich, Otto-Stern-Weg 1, 8093 Zürich, Switzerland}
\newcommand{\affF}{Quantum Center, ETH Zurich, CH-8093, Switzerland}
\newcommand{\affE}{ZuriQ AG, Otto-Stern-Weg 1, 8093 Zürich, Switzerland}

\begin{document}

\preprint{APS/123-QED}

\title{A magnetic-field insensitive gate set for trapped-ion nuclear spin qubits}%

\author{J. P. Home}
\email{jhome@phys.ethz.ch}
\affiliation{\affD}\affiliation{\affF}
\author{J. Flannery}
\affiliation{\affD}\affiliation{\affF}\affiliation{\affE}
\author{M. Mazzanti}
\affiliation{\affD}\affiliation{\affF}
\author{J. Apolin}
\affiliation{\affD}\affiliation{\affF}
\author{S. Jain}
\affiliation{\affE}

\date{\today}

\begin{abstract}
We outline a complete set of operations for manipulating nuclear spin qubits of trapped-ions stored in high magnetic fields such as required for Penning trapping, where the nuclear and electron spins are largely decoupled due to the dominance of the external field Hamiltonian. The reduced nuclear magnetic moment results in insensitivity to external magnetic fields compared to the use of an electron spin, but also makes the nuclear spin hard to manipulate on fast timescales. To maintain speed, we propose a two-qubit phase gate technique which utilizes the electron spin, but for which the qubit remains protected from magnetic fields through retaining the nuclear encoding. This method works for magnetic-field insensitive qubits at lower fields as well as for both microwave and laser field gradients. 
\end{abstract}

\maketitle

\section{Introduction}

The use of Penning traps for quantum computing offers several potential advantages vs. radio-frequency trapping, due to the natural capability for moving ions in 3-dimensions \cite{jainPenningMicrotrapQuantum2024} and the lack of need to apply high radio-frequency voltages to electrodes, which aids in design and fabrication of ion trap chips \cite{jainScalableArraysMicroPenning2020}. One challenge however is the presence of the large magnetic field, typically on the order of several Tesla \cite{jainPenningMicrotrapQuantum2024, mcmahonSecondScaleBeSpin2022, brittonEngineeredTwodimensionalIsing2012, shigaDiamagneticCorrectionBe2011}, which largely decouples the nuclear and electron spin, since the magnetic Hamiltonian dominates over the hyperfine interaction. The electron spin energy levels have a high sensitivity to these external field fluctuations, given approximately by $g_s \mu_B B$ which corresponds to a qubit frequency dependence of 28~GHz/T. Using these to store qubits leaves the challenge of stabilizing the large magnetic field to levels of $\delta B/B \sim 10^{-10}$ in order to achieve coherence times over 10~ms \cite{brittonVibrationinducedFieldFluctuations2016}. While at lower intermediate magnetic field values first-order field-independent points for electronic transitions exist at which high insensitivity to the magnetic field strength can be attained, at the fields required to produce megahertz trapping frequencies in Penning traps no such transitions are available. An alternative encoding for a qubit would be to use the nuclear spin, where the sensitivity to magnetic fields is reduced by approximately $g_I \sim m_e/m_p \sim 1/1800$ (a more careful analysis is given below). However the same reduction which aids in qubit coherence results in control of the nuclear spin being significantly reduced in speed. While these reduced matrix elements can be overcome by strong driving, it is desirable to minimize the need for this. This challenge is particularly stark for two-qubit gates between ions, for which the relevant Rabi frequencies are also reduced by the ratio of the zero-point motion size to the length scale of the gradient of microwave or optical fields (for the latter, this ratio is given by the Lamb-Dicke parameter).

In this article, we work through the relevant considerations for performing a complete two-qubit gate set on nuclear spin qubits, considering implementation using microwave fields. We first solve for the atomic physics, and give the relative rates and energy levels at the fields involved, comparing the electron and nuclear spin single-qubit control and coherence properties. We then consider how best to perform microwave two-qubit gates in such a system, trying to avoid compounding the reduced speed of nuclear transitions with further speed reductions due to the relative weakness of field gradients. To do this we consider producing phase gates using an oscillating field gradient which provides a state-dependent potential for the electron spin. We show that the combination of a 4-tone drive close to resonance with electron spin-flip transitions with a low-frequency state-dependent magnetic field gradient provides a gate in which the qubit can remain protected from noisy external magnetic fields while retaining the speed of gates performed directly on the electron spin. This provides a promising avenue enhancing the prospects of Penning traps for quantum computation, through the combination of robust qubit encodings with high fidelity gates. 

\section{Energy levels}
The Hamiltonian for the hyperfine ground state of an alkali-like ion in a magnetic field $\bm B$ is given by 
$H = H_{\rm hfs} + H_B$ where 
\be
H_{\rm hfs} = A{\bf I}\cdot \bm{\sigma} = A \left(I_z \sigma_z + I_x \sigma_x + I_y \sigma_y\right)
\ee gives the hyperfine interaction, and 
\be 
H_B = (g_S \mu_B \bm{\sigma} + g_I \mu_B {\bf I } )\cdot{\bf B}
\ee  gives the interaction with the magnetic field. Here $\bm{\sigma}\equiv (\sigma_x, \sigma_y, \sigma_z)$ is the electron spin operator and ${\bf I} \equiv (I_x, I_y, I_z)$ that of the nuclear spin, and $A$ is the hyperfine constant for the atom in question. We choose the magnetic field along the $z$ axis.  In conditions used for Penning traps, the high magnetic field presents a hierarchy of energy scales such that  $|g_S \mu_B B/h| \sim 100\, {\rm GHz} \gg |A|/h \sim 1-3\ {\rm GHz} \gg |g_I \mu_B B/h| \sim 10\ {\rm MHz}$ (throughout we use a convention in which the scaling of the nuclear coupling to magnetic fields is incorporated into the g-factor, ie. $g_I \sim m_e/m_p g_S$). This means that the eigenstates of the total Hamiltonian $H = H_{\rm hfs} + H_{B}$ are primarily defined by $H_B$, and are thus close to the eigenstates of $\sigma_z, I_z$. Making a lowest order approximation which ignores terms with higher order dependence on $A$, the energies are given by
\begin{equation}
	E_{M, M_s} = g_S \mu_B M_s B + (A M_s + g_I \mu_B B)  M 
\end{equation}
with $M, M_s$ the eigenvalues of the $I_z, \sigma_z$ operators respectively. The off-diagonal terms of the hyperfine interaction then can be treated as perturbations to this simple structure. The combination of $\sigma_x I_x + \sigma_y I_y $ can be written in terms of the raising and lowering operators (defined as $\sigma_\pm = \sigma_x \pm i \sigma_y$ and $I_\pm = I_x \pm i I_y$) for the relevant angular momentum components to give $A/2 (\sigma_+ I_- + \sigma_- I_+)$. This couples levels with the same $z$ projection of total angular momentum ($M_S + M$) which differ in frequency by an amount which is on the order of the electron spin splitting.  For $M < I$, these span the subspace $\ket{\uparrow}_{M} \equiv |M_s = 1/2, M \rangle, \ket{\downarrow}_M \equiv |M_s = -1/2, M+1\rangle $, which have mean energy $\bar{E}_{M} = -A/4 + g_I \mu_B  B (M + 1/2)$. In this subspace, the Hamiltonian can be written as  
\be
H_{M} = \bar{E}_{M} + \Delta_{M} \sigma_Z^{M}  + g_{M} \sigma_X^{M}
\ee
where $\Delta_{M}  \simeq g_s \mu_B B/2 + A M/2 + A/4 $, $g_I \mu_B B/2 \ll g_s \mu_B B/2$, $g_{M} =\frac{A}{2}\sqrt{I(I + 1) - M (M +1)}$ and the Pauli operators are defined as 
\be 
\sigma_Z^{M} &\equiv& \ket{\uparrow}_{M}\bra{\uparrow}_{M} - \ket{\downarrow}_{M}\bra{\downarrow}_{M}  \\
\sigma_X^{M} &\equiv& \ket{\uparrow}_{M}\bra{\downarrow}_{M} +  \ket{\downarrow}_{M}\bra{\uparrow}_{M} \ .
\ee 
The eigenvalues of this system are well known, giving  the energies
\be
E_{M, \pm 1/2}  = \bar{E}_{M} \pm (\Delta_{M}  ^2 + g_{M}^2)^{1/2}\\ \simeq \bar{E}_{M} \pm \Delta_{M}  \pm \frac{g_{M}^2 }{2 \Delta_{M} } \ .
\ee
The eigenvectors are approximated by 
\be
\ket{+}_{M} &\simeq& \ket{1/2, M}  + \frac{g_{M}}{\Delta_{M} }\ket{-1/2, M + 1}\\
 \ket{-}_{M} &\simeq& - \ket{-1/2, M + 1} + \frac{g_{M}}{\Delta_{M} }\ket{1/2, M} \ .
\ee
An important ratio in all of these expressions is $g_M/\Delta_M$ which is of order $1/100$ for typical fields used in experiments (a summary of real values for calcium and beryllium experiments is given below). Here we see that the hyperfine interaction mixes in a small fraction of the opposite spin state (and $M$ shifted by 1). This has the effect of allowing oscillating magnetic fields to be used to couple the $M$ manifold as defined above to other manifolds, thus enhancing the possibility to use a nuclear spin qubit to store information, without having to directly drive nuclear spin flips with the drive fields themselves.

\section{Qubit choices}
\begin{figure}[t]
    \centering
    \includegraphics[width=1.0\columnwidth]{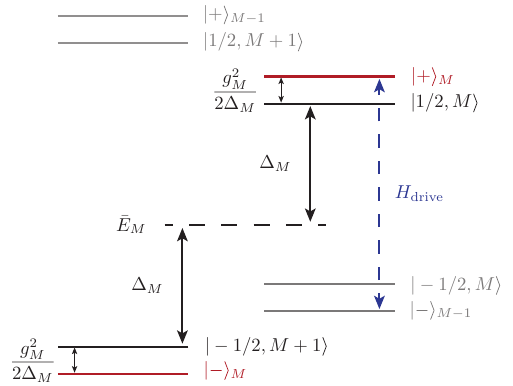}
    \caption{Relevant energy levels and relative splittings.}
    \label{fig:gsstructure}
\end{figure}

The energy levels for such a system are shown in figure \ref{fig:gsstructure}. For the definition of qubits, the sensitivity of the difference of energy levels to changes in the magnetic field are important. For pairs of levels separated by an electron spin flip - we will refer to these as ``electron'' qubits, this is dominated by the energy shifts through coupling of the electron spin to the magnetic field, giving a field dependence of $d \omega/d B \sim g_s \mu_B/\hbar ~ \sim 2 \pi \times 28$~GHz/T. In fields of several Tesla, achieving coherence times on the order of a second corresponds to stabilizing the magnetic field to 1 part in $10^{12}$, which is highly challenging. On the other hand, transitions between levels in an electron spin qubit can be driven fairly efficiently due to the strong coupling to electro-magnetic fields.

An improved insensitivity to the strength of the quantization field can be achieved through defining a qubit between two ``nuclear'' levels, ie. a qubit choice of $\{\ket{-}_{M}, \ket{-}_{M - 1}\}$, which in the limit of vanishing hyperfine mixing would be nuclear spin states with the same electron spin. For this choice the sensitivity of the transition energy to changes in the magnetic field is 
\be
\frac{d(\Delta E)}{dB} \simeq g_I \mu_B - \frac{A^2}{2 B^2 g_s^2 \mu_B^2} g_s \mu_B M\ .
\ee
where a number of smaller contributions have been neglected. Taken as a ratio to the susceptibility of the bare electron spin $g_s \mu_B$ we get 
\be
\chi_{\rm rel} \simeq \frac{g_I}{g_S} - \frac{A^2}{2 B^2 g_s^2 \mu_B^2} M\ .
\ee
In the absence of hyperfine mixing this reduces to the first term, corresponding to the nuclear spin sensitivity - a  factor of around 1/1800 reduced vs. the electron spin. The second term corresponds to a factor of around $10^{-4}$ for calcium and beryllium ions at typical working fields of a few Tesla. Thus we see that at these fields the first term would  be expected to dominate. Penning trap systems have been able to demonstrate coherence for electron spins up to 100~ms \cite{shigaDiamagneticCorrectionBe2011}; an extension by a factor of 1800 would correspond to 180~seconds.

\section{Single qubit gates}
Single qubit gates can be driven in atomic systems either by near-resonant optical or microwave fields. Here we consider microwave driving, which is relevant for the energy scales of the ground state structure. Transitions are driven by oscillating magnetic fields, $B_d$, which couple to both the electron and nuclear spin. Since the field couples much more strongly to the electron spin, its effect dominates the resulting dynamics. Making the approximation that only electron spin flips are driven, the Hamiltonian for the drive can be written as
\be
H_{\rm drive} = g_s \mu_B B_d \cos(\omega t) \sigma_x
\ee
with $\sigma_x \equiv \left(\ket{1/2}\bra{-1/2} + \ket{-1/2}\bra{1/2} \right)\otimes I_M$ with $I_M$ the identity operation on the nuclear spin.
Choosing the frequency $\omega$ to be close to resonance allows to make a rotating wave approximation. Since the microwave drive Hamiltonian couples states which have the same nuclear spin quantum number $M$ but different electron spin states, the relevant pairs of states are  $\{\ket{+}_{M}, \ket{+}_{M - 1}\}$, $\{\ket{-}_{M}, \ket{-}_{M - 1}\}$ and $\{\ket{+}_{M}, \ket{-}_{M - 1}\}$. The strength of the couplings in these cases are given by the matrix elements, which are 
\be
\bra{+}_{M} \sigma_x\ket{+}_{M - 1} &\simeq \frac{g_{M-1}}{\Delta_{M - 1}} \\ 
\bra{-}_{M} \sigma_x \ket{-}_{M - 1} &\simeq \frac{g_{M}}{\Delta_{M}}\\ 
\bra{+}_{M} \sigma_x \ket{-}_{M - 1} &\simeq 1  \ .
\ee
We see that the Rabi oscillations for transitions on the nuclear qubits are a factor $g_{M - 1}/\Delta_{M-1} \sim 1/100$ lower than that for a direct electron spin flip transition for typical working conditions. Thus driving the  nuclear transition is either slow or requires more significant magnetic field drives, corresponding to higher currents in eg. microwave drive lines. It is notable that the improvement of coherence time is larger than this expected reduction in Rabi rate. However for two-qubit gates, which are typically slower than single-qubit gates due to the need to utilize field gradients to produce forces, the slow-down associated with driving nuclear spin flips is undesirable. Thus the two-qubit gates described below aim to avoid this, and rather produce conditional phase shifts between these states rather than nuclear spin flips.

\section{Experimental parameters}
To summarize the results of the atomic physics discussed above, Table \ref{tab:expparams} gives example parameters for light ions which seem most attractive for quantum computing in Penning traps due to their low mass and consequentially higher bare cyclotron frequency, which governs the ion oscillation frequencies which can be achieved (axial frequencies up to $\omega_c/\sqrt{2}$ produce stable trapping). We base these numbers on two magnetic field values, 3~T and 6~T, to show the variation which might be expected. The most relevant ions for Penning traps working at these fields are beryllium and calcium, where in both cases the odd-isotope allows for the use of the nuclear spin. To get an idea of achievable speeds we take values realized in recent experiments on high fidelity gates in radio-frequency traps. For microwave fields near $100~\mathrm{GHz}$, we expect single qubit Rabi frequencies of $50$–$100~\mathrm{kHz}$ to be easily achievable with waveguide delivery. Higher rates could be achieved by integrating microwave resonators into the trap.

\begin{table}[ht]
    \centering
    \begin{tabular}{c|c|c|c|c|c|c}
        Species & $B$ & $\left|\frac{g_{M}}{\Delta_{M}}\right|$ & $\chi_{\rm rel}$ & $\omega_c/(2 \pi)$ & $\omega_e/(2 \pi)$ & $\omega_n/(2 \pi)$ \\
       &  (T)  & & ($\times 10^{-4}$) & (MHz) & (GHz) & (MHz) \\ \hline 
       $^9$Be$^+$  & 3 & 0.013 & 1.7 & 5.11 & 83.2 & -298 \\
         $^9$Be$^+$ & 6 & 0.006 & 2.0  & 10.2 & 167.2 & -278 \\
           $^{43}$Ca$^+$  & 3 & 0.026 & 0.16 &  1.07 & 81.3 & -399 \\
         $^{43}$Ca$^+$ & 6 & 0.013 & 1.6 & 2.14 & 165.3 & -376
    \end{tabular}
    \caption{Values for quantities relevant to experimental setups for Penning traps. Results are quoted at magnetic fields of 6 and 3~Tesla. Values for atomic constants used to calculate these numbers are taken from \cite{tiesingaCODATARecommendedValues2021, nortershauserPrecisionTestManyBody2015,bollingerLasercooledStoredIon1985, shigaDiamagneticCorrectionBe2011}.}
    \label{tab:expparams}
\end{table}

\section{Two-qubit gate}
The relative immunity of the nuclear encodings to fluctuations of the magnetic field makes them attractive for quantum logic, but the relatively weak coupling for driving the nuclear qubit raises the question how to perform multi-qubit gates. Multi-qubit gates performed with trapped ions are anyway slower than single-qubit gates due to the need for spin-motion couplings which rely on a spatial gradient of the driving field. While for laser-based schemes achieving suitable linearity of the coupling requires a slow-down factor of $0.1$ \cite{benhelmFaulttolerantQuantumComputing2008, ransfordHelios98qubitTrappedion2025}, for direct magnetic field drives, gradients are harder to achieve, with the use of near-field effects predominating \cite{ospelkausTrappedIonQuantumLogic2008, hughesTrappedionTwoqubitGates2025}. In order not to lose in speed from both the need for nuclear spin driving \emph{and} the need for motional gradients, it is reasonable to consider how gates could be performed by driving the electron spin, and whether this is possible without compromising the insensitivity to magnetic field fluctuations. Furthermore it seems more difficult to generate field gradients at frequencies of $100$~GHz close to the electron spin flip than at either the nuclear spin flip frequencies in the 100~MHz range or motional frequencies in the few MHz range. Thus we restrict ourselves to gates where the field gradients for state-dependent forces are in these lower frequency ranges. We show that all of these  requirements can be satisfied using a method (section \ref{sec:4tonegate}) which modifies schemes recently realized in radio-frequency traps \cite{srinivasHighfidelityLaserfreeUniversal2021a, sutherlandLaserfreeTrappedionEntangling2020}. 

\subsection{Reduced notation}
\begin{figure}[t]
    \centering
    \includegraphics[width=\columnwidth]{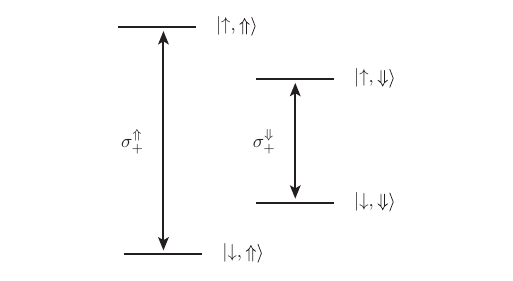}
    \caption{Simplified level definition used for the discussion of two-qubit gates. The electron spin-flip operators for each transition are labelled as $\sigma_\pm^i$with superscripts $i = \Uparrow, \Downarrow$.}
    \label{fig:simplified}
\end{figure}

For discussion of two-qubit gates, the notation above, which carries all the relevant quantum numbers, is cumbersome. Hence we will focus on the generic 4-level system shown in figure \ref{fig:simplified}. The notation used is that of the corresponding states in the limit that the hyperfine mixing is tuned to zero. We only consider two nuclear spin values, hence we choose a simplified notation in which these can take values $\Uparrow$ and $\Downarrow$, while the electron spin is given by $\uparrow$, $\downarrow$. Given the large electron spin energy scale, the lowest two levels are thus $\ket{\downarrow, \Downarrow} \equiv \ket{-}_{M - 1}, \ket{\downarrow, \Uparrow}  \equiv \ket{-}_M$ and the upper levels $\ket{\uparrow, \Downarrow} \equiv \ket{+}_{M - 1}, \ket{\uparrow, \Uparrow} \equiv \ket{+}_M$, where these identifications are valid for $M>1/2$. Transitions in the following sections are assumed to drive the electron spin transitions while conserving the nuclear spin. Hence there are two transitions, which we will label with the nuclear spin value. Operators for the electron spin flip will be denoted as $\sigma_z, \sigma_\pm$. These hold regardless of the nuclear spin state. However it is sometimes convenient to separate out laser fields resonant with the transitions relevant for one nuclear spin state from the other. To do this, we introduce an additional notation including a projector onto the relevant nuclear state:   $\sigma_z^n, \sigma_\pm^n$ with $n\in \{\Uparrow, \Downarrow\}$, ie. $\sigma_+^\Uparrow \equiv \sigma_+ \otimes \ket{\Uparrow}\bra{\Uparrow}$ denotes an electron spin-flip operator, driven by fields which are near resonant with the transitions of the $\Uparrow$ state. 

The energy scales of the different types of transitions are diverse. Driving an electron spin flip requires microwave frequencies in the 100~GHz range, where it seems challenging to achieve high Rabi frequencies. Nuclear spin flips in the few 100~MHz--1~GHz range have much smaller matrix elements, but it should be easier to achieve the high currents required to drive these. Meanwhile the motional frequencies, which we couple to using electron-state dependent potentials are typically on the MHz level \cite{jainPenningMicrotrapQuantum2024}.

\subsection{Direct phase gate}
\label{sec:basicgate}
One of the traditional two-qubit gates used in trapped ions is a phase gate \cite{leibfriedExperimentalDemonstrationRobust2003, homeDeterministicEntanglementTomography2006}, resulting from a Hamiltonian 
\be
H_Z = 2 \hbar \Omega_g \cos(\omega_g t) S_z a e^{-i \omega_z t} + {\rm h.c.} 
\ee
where $S_z = \sum_{j = 1}^2 \sigma_{z,j}$, $\omega_z$ is the frequency of one of the collective modes of oscillation of the two ions with lowering operator $a$, and $\Omega_g$ gives the coupling between the spin and motion. The Hamiltonian takes the form of a force which is dependent on the internal states of the ions. Here it is assumed that the spin-dependent drive is in-phase on both ions, and that the motional mode is common mode (the generalization to unequal modes and driving are straightforward \cite{zhuTrappedIonQuantum2006, homeCompleteMethodsSet2009, homeChapter4Quantum2013}). Such a Hamiltonian can be produced through the use of optical dipole potential or magnetic field gradients. When the drive is close to the motional mode frequency $\epsilon \equiv \omega_g - \omega_z \ll \omega_z$, we can make a rotating wave approximation, resulting in 
\be
\label{eq:GateForce}
H_{Z,I} = \hbar \Omega_g S_z \left\{a e^{i \epsilon t} + {\rm h.c.} \right\}
\ee
which can be considered as a state-dependent force with force strength $F \propto 2 \hbar \Omega_g$.
Exponentiation of this (time-dependent) Hamiltonian leads to a unitary evolution 
\be
U(t) = D(\alpha(t) S_z) e^{i \Phi(t) S_z^2} 
\ee
with $\alpha(t) = 2 i e^{-i \epsilon t/2} (\Omega_g/\epsilon) \sin(\epsilon t/2)$, the internal state dependent displacement $D(\alpha) = \exp(\alpha^* a -  \alpha a^\dagger)$ and the phase 
\be
\Phi(t) = \frac{\Omega_g^2}{\epsilon^2} \left(\sin(\epsilon t) - \epsilon t \right) \ .
\ee
At $N$ multiples of the revival time $t_r = 2\pi/\epsilon$, the displacement vanishes and the resulting accumulated phase is 
\be
\Phi(N t) = N \Phi_0 = 2 \pi N  \frac{\Omega_g^2}{ \epsilon^2} \ .
\ee
To produce a two-qubit gate equivalent to the phase gate in a single revival time ($N = 1$) requires $\Omega_g = \epsilon/4$, for which $U(t_r) = {\rm Diag}\{1, i, i, 1\}$.

Suitable gradients can be produced either using AC Stark shifts using travelling standing waves with spatially varying polarization \cite{leibfriedExperimentalDemonstrationRobust2003, homeDeterministicEntanglementTomography2006} or intensity \cite{clarkHighFidelityBellStatePreparation2021}. Alternatively it is possible to use an oscillating magnetic field gradient $\partial B_g/\partial \alpha$ where $\alpha$ represents a spatial direction ($x,y,z$) which must have a component along the relevant motional mode. Such gradients which can be generated using currents passing through the surface of the chip. For a field gradient along the $z$ axis we find
\be
\Omega_g = z_0 g_s \mu_B \frac{\partial B_g}{\partial z} \ .
\ee
with $z_0$ the zero point extent of motion of the oscillator mode utilized for the gate \cite{ospelkausTrappedIonQuantumLogic2008}. For optical gradients the factor $z_0 \partial B_g/\partial z$ is replaced by the Lamb-Dicke parameter $\eta = k z_0$, with $k$ a relevant wavevector for the light field which is determined by the precise experimental setup. 

While the phase gate has been used extensively for optical fields \cite{leibfriedExperimentalDemonstrationRobust2003, homeDeterministicEntanglementTomography2006, clarkHighFidelityBellStatePreparation2021}, for the magnetic gradient implementation it implies currents applied to the electrodes of the trap at close to the motional frequencies. It is difficult to avoid these currents being accompanied by oscillating voltages, which then leads to driving of the ion by state-independent electric fields, leading to large single qubit phases \cite{ospelkausTrappedIonQuantumLogic2008}. For this reason the direct phase gate has not been used to perform high-fidelity gates, but rather an adapted scheme has been used which is described below \cite{srinivasHighfidelityLaserfreeUniversal2021}. 

\subsection{Microwave enhanced phase gate}
In order to avoid the adverse effects of voltages which accompany oscillating currents close to the trap frequency, Sutherland et al. \cite{sutherlandLaserfreeTrappedionEntangling2020} proposed detuning the oscillating field gradient from the motional frequency of the ions, but then using two additional microwave tones with strength $\Omega_\mu$ which are detuned from the qubit transition frequency by $\pm \delta$, where $\delta$ bridges the frequency difference between the gradient drive and the trap frequency. This is similar to schemes originally proposed and implemented for static magnetic field gradients \cite{mintertIonTrapQuantumLogic2001, weidtTrappedIonQuantumLogic2016}. The Hamiltonian can be written in the interaction picture with respect to the internal state energy levels separated by $\hbar \omega_0$ and the motion at frequency $\omega_z$, keeping only resonant terms with respect to $\omega_0$  \cite{sutherlandLaserfreeTrappedionEntangling2020}
\be
\label{eq:twotone}
H(t) &=& 2 \hbar \Omega_\mu S_+ \cos(\delta t + \phi) + {\rm h.c.} \nonumber \\ 
&+& 2 \hbar \Omega_g \cos(\omega_g t) S_z a e^{-i \omega_z t} + {\rm h.c.} \ .
\ee 
Compared to the work of \cite{sutherlandLaserfreeTrappedionEntangling2020}, this expression differs only in the inclusion of the phase of the modulation $\phi$, which will be important for what follows. A judicious choice of interaction picture, transformed according to 
\be
U_\mu(t) = \exp{\left[ - i 2 S_x \Omega_{\mu} \sin(\delta t + \phi)/\delta\right]}
\ee
gives rise to the interaction Hamiltonian 
\be
\label{eq:fullinteraction}
H_I(t) &=& 2 \hbar \Omega_g \cos(\omega_g t) \left\{ a e^{-i \omega_z t } + a^\dagger e^{i\omega_z t}\right\} . \\ \nonumber & & \left[ S_z \left\{J_0(\beta)  + 2 \sum_{n = 1}^\infty J_{2 n}(\beta) \cos\bigl(2 n (\delta t + \phi)\bigr)  \right\} \right. \\  \nonumber &+& \left. 2 S_y \sum_{n = 1}^\infty J_{2 n -1 }(\beta) \sin\bigl((2 n -1 )(\delta t + \phi)\bigr) \right]
\ee
where $\beta = 4 \Omega_\mu/\delta$ and the $J_{k}(\beta) $ are Bessel functions of the first kind. Following the original proposal \cite{sutherlandLaserfreeTrappedionEntangling2020}, choosing $2 \delta = \omega_z - \omega_g + \epsilon $ allows to pick out a near-resonant drive which approximates the state-dependent force Hamiltonian
\be
H_F(t) = \hbar \Omega_g J_{2}(\beta) S_z a e^{i (\epsilon t + 2 \phi)} + {\rm h.c.}  \ .
\ee
We see that the phase of the force can be controlled using the relative phases of the two microwave tones. This feature will be used in section \ref{sec:4tonegate} below.  Since this Hamiltonian is the same as that of equation \eqref{eq:GateForce}, the same methods can then be used to perform a two-qubit gate. The phase acquired is modified by the Bessel function squared 
\be
\Phi(N t) = N \Phi_0 = 2 \pi N  \frac{\Omega_g^2 J_s(\beta)^2}{\epsilon^2} \ .
\ee
which when set to $\pi/2$ gives rise to the same phase gate operation as previously, modifying $\epsilon$ or $\Omega_g$ to accommodate the modification due to the Bessel function. Such a gate could be performed directly on a qubit involving an electron spin-flip, such as $\ket{\uparrow \Uparrow},\ket{\downarrow \Uparrow}$ for which the energy separation is sensitive to the magnetic field. Judicious choices of $\beta$ can lead to protection from magnetic field noise affecting the qubit transition, which has been proposed \cite{sutherlandLaserfreeTrappedionEntangling2020} and demonstrated in experiments \cite{srinivasHighfidelityLaserfreeUniversal2021a}. However for a nuclear encoded qubit, the force on both levels is not strongly state-dependent. This leads to the consideration of how ancilliary electron spin-flip transitions could be used to perform the two-qubit gate. 

\section{Gates on a nuclear qubit}
\subsection{Direct approaches through qubit drives}
Since nuclear qubits have weak dependence on external optical and magnetic fields, the methods of the previous sections become impractical when  applied directly. One approach to overcoming this problem would be to map qubit information between nuclear storage and the electron spin \cite{homeCompleteMethodsSet2009}. For instance a nuclear qubit $\ket{\Downarrow, \downarrow}, \ket{\Uparrow, \downarrow}$ could be driven using a transfer pulse resonant with the $\Downarrow$ electron spin-flip transition ($\ket{\Downarrow, \downarrow} \leftrightarrow \ket{\Downarrow, \uparrow}$ ), mapping the encoding to $\ket{\Downarrow, \uparrow}, \ket{\Uparrow, \downarrow}$. While this would suffice for the direct phase gate, the qubit encoding in the electron spin is susceptible to magnetic field noise. The  microwave enhanced phase gate is not possible in this approach since it requires to directly drive the qubit transition, and in this case the matrix element is small. 

\begin{figure}[t]
    \centering
    \includegraphics[width=0.9\columnwidth]{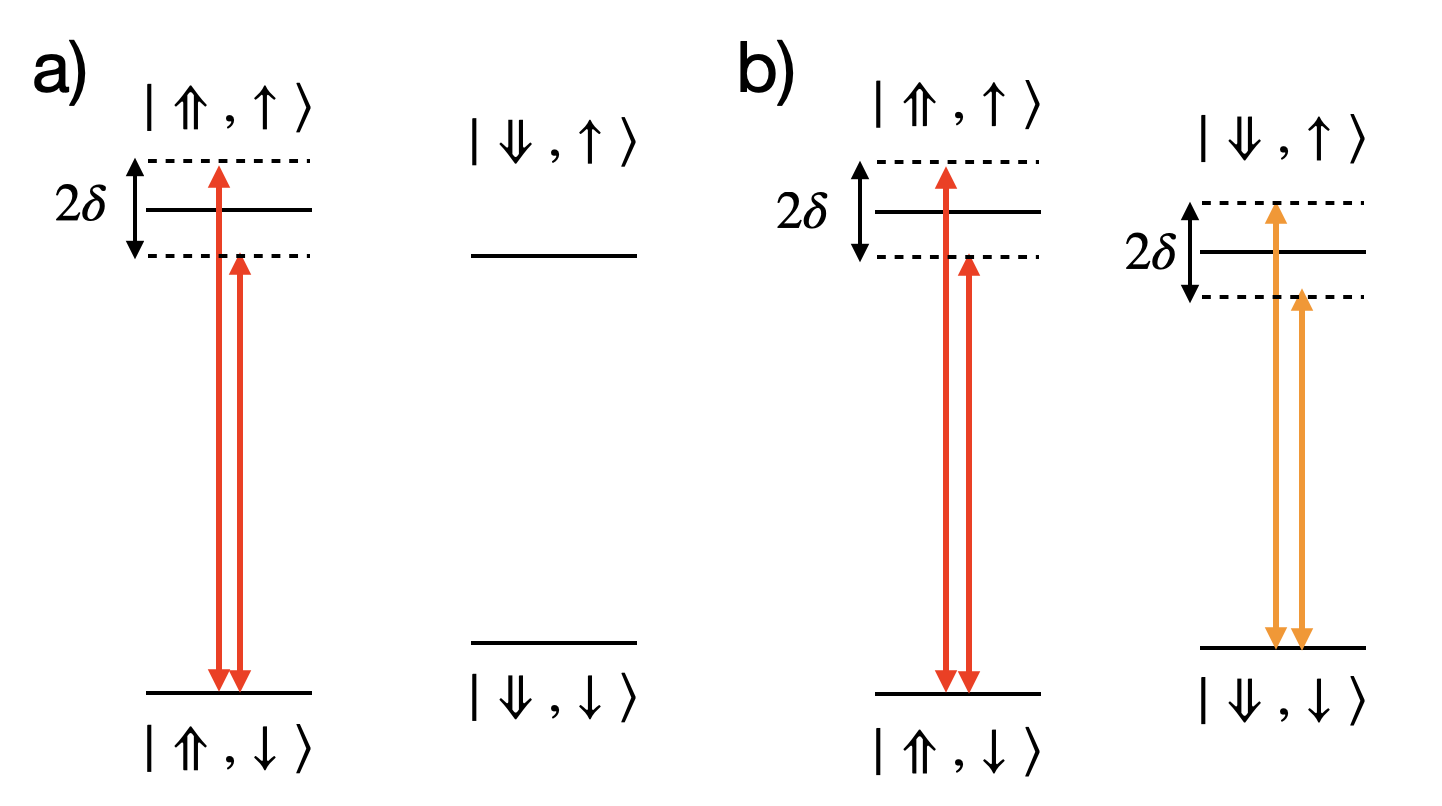}
    \caption{Two qubit gate schemes. a) The two-tone scheme involves microwave dressing fields applied $\pm \delta$ detuned from the $\Uparrow$ electron spin flip transition. This gate can be used to perform a phase gate on the nuclear spin, but the atom leaves the pure nuclear spin encoding during the gate. b) The four-tone gate uses drives on both electron spin-flip transitions simultaneously, with the same strength. The only difference is that the two tones driving the $\Downarrow$ transition have a relative phase which is offset from that of the two tones for the $\Uparrow$ transition $\pi/2$. This ensures that the Hamiltonian acting on the nuclear spin states has a force direction which depends on the nuclear spin, thus allowing a gate directly on the nuclear qubit. }
    \label{fig:gatescheme}
\end{figure}

A scheme which does allow the microwave-enhanced method to be used is to keep the qubit encoded in $\ket{\Downarrow, \downarrow}, \ket{\Uparrow, \downarrow}$, but then perform the gate directly by implementing a two-tone drive Hamiltonian of equation \ref{eq:twotone} directly on the $\Uparrow$ electron spin-flip transition $\uparrow \leftrightarrow \downarrow$ (illustrated in figure \ref{fig:gatescheme} a)). The $\ket{\Downarrow \downarrow}$ state is then a spectator to the dynamics. In this form the action of the Hamiltonian on the qubit subspace is 
\be
H_F(t) = \hbar \Omega_g J_2(\beta) (I_4 + S_z^{\rm nuc}/2) \left\{a e^{i \epsilon t + 2 \phi} + {\rm h.c.}\right\}
\ee
where $S_z^{\rm nuc} = \sigma_z^{\rm nuc}\otimes I + I\otimes \sigma_z^{\rm nuc}$ and $\sigma_z^{\rm nuc} = \ket{\Downarrow}\bra{\Downarrow}-\ket{\Uparrow}\bra{\Uparrow}$. The square of the spin operator produces the operation
\be
(I_4 + S_z^{\rm nuc}/2)^2 = I_4^2 + S_z^{\rm nuc} + \frac{1}{4}\left(S_z^{\rm nuc}\right)^2 \ .
\ee
The last term on the right hand side generates the relevant phase for the two-qubit gate. Since this is 4 times smaller than for the direct gate, the target phase $\Phi$ has to be 4 times larger. For the same microwave power, compensation can be achieved by increasing $N$ or decreasing $\delta$. This corresponds to a factor of 2 in slow-down in gate if only the detuning is used, but comes at the cost of higher spin-motion entanglement and thus a higher susceptibility to motional decoherence errors. Increasing the number of loops alone requires 4 times as long a gate, thus the gate is more susceptible to a variety of decoherence channels. In this implementation the gate is also susceptible to dephasing due to fluctuations in the magnetic field, since the qubit spends a significant time with the electron spin inverted - as will be seen in section \ref{sec:magfieldsusc}, this cannot be significantly suppressed through judicious choice of $\Omega_\mu$, as is possible if such a gate is performed directly on an electron spin qubit. For this reason, in the next section we consider a modified scheme which both reduces the gate time and greatly suppresses the sensitivity to magnetic field fluctuations.

\subsection{Four-tone gate}
\label{sec:4tonegate}
As an alternative to performing the gate using forces which apply to only one of the nuclear spin states, here we instead consider using 4 microwave tones, two of which are tuned to each of the relevant electron spin transitions, to drive both of the nuclear spin states simultaneously. This setup is illustrated in figure \ref{fig:gatescheme} b). A critical element of this alternative is that the two tones driving each of the nuclear spin states have a controllable relative phase, which can be used to set the sign of the force that acts on the respective state. Thus with a single oscillating gradient which produces a state-dependent shift of the electron spin which is independent of the nuclear spin, we have 
\be
H(t) &=& \sum_{i = \Uparrow, \Downarrow} 2 \hbar \Omega_{\mu, i} S_{+}^i \cos(\delta_i t + \phi_i) + {\rm h.c.} \nonumber \\ &+& 2 \hbar \Omega_g \cos(\omega_g t) \sum_{i = \Uparrow, \Downarrow} S_{z}^i a e^{-i \omega_z t} + {\rm h.c.}
\ee
where the index $i = \Uparrow, \Downarrow$ is used to label the transition under consideration. The amplitude, relative phase and relative detunings of the drives on each transition are independently controllable. In a slight modification to the 2-level case, we use a unitary transformation 
\be
U(t) = \exp{\left[ - 2 i  \sum_{i = \Uparrow, \Downarrow} S_{x}^i \Omega_{\mu, i} \sin(\delta_i t + \phi_i)/\delta_i\right]} 
\ee
and choosing $\Omega_{\mu, \Uparrow} = \Omega_{\mu, \Downarrow}$ and both $2 n \delta_i = \omega_z - \omega_g + \epsilon$ gives a state-dependent force Hamiltonian for the four-level system 
\be
H_F(t) = \hbar \Omega_g J_{2}(\beta) \sum_{i = \Uparrow, \Downarrow} S_{z}^i \left\{ a e^{-i \epsilon t - i 2 \phi_i} + a^\dagger e^{i\epsilon t + i 2 \phi_i}\right\} \ .
\ee
Let us now consider a nuclear qubit $\ket{\downarrow \Downarrow}, \ket{\downarrow \Uparrow}$ which corresponds to the lower energy level of both transitions.  If the two relative phases $\phi_i$ are equal, these two states will experience the same force, which precludes their use in a two-qubit gate. However it is also possible experimentally to set $\phi_1 = \phi_2 + \pi/2$, which will lead to each state experiencing an opposing force. Restricting ourselves to the terms relevant to these states through the projector $P_N = \ket{\downarrow \Downarrow}\bra{\downarrow \Downarrow} + \ket{\downarrow \Uparrow}\bra{\downarrow \Uparrow}$, the force Hamiltonian becomes
\be
P_N H_F = \hbar \Omega_g  J_{2}(\beta)  S_z^{\rm nuc} \left\{ a e^{-i \epsilon t} + a^\dagger e^{i\epsilon t}\right\} \ .
\ee
This is the Hamiltonian of equation \ref{eq:GateForce}, but with the nuclear spin operator replacing the electron spin. A gate can thus be performed in an analogous manner to section \ref{sec:basicgate}, with identical duration. 

\section{Susceptibility to magnetic fields}
\label{sec:magfieldsusc}
The gates described in the previous sections provide differing implementations for a nuclear qubit in a high-field system such as a Penning trap. However they differ significantly in their susceptibility to magnetic field fluctuations. To lowest order, the interaction with a magnetic field can be modeled as coupling to the electron spin. Taking for simplicity a model in which the noise is at a single frequency $\omega_N$ and has an amplitude $g_N$, we can write a single frequency oscillating Hamiltonian 
\be
H_N = \hbar g_N \cos(\omega_N t) \left( S_{z, \Uparrow} + S_{z, \Downarrow}\right) \ .
\ee
As shown previously \cite{sutherlandLaserfreeTrappedionEntangling2020}, the effect of the modulating microwave tones is to multiply this term by a Bessel function. For an electron spin qubit, this allows one to choose the modulation such that $J_0(\beta) = 0$ to suppress the sensitivity to noise at frequencies $\omega_N < \delta$ in a technique called intrinsic dynamical decoupling (IDD). 

For the gates on the nuclear qubit, the first scheme with just a single set of microwave drives on the $\Uparrow$ transition, the result of the modulation is to transform the noise Hamiltonian to  
\be
H_N^{(2)} = \hbar g_N \cos(\omega_N t)  \left( S_{z, \Downarrow} + J_0(\beta) S_{z, \Uparrow}\right) \ .
\ee
Thus a sensitivity to magnetic field fluctuations at a similar order of magnitude to that of an electron spin is introduced between the two nuclear spin states (\ket{\downarrow, \Uparrow}, \ket{\downarrow, \Downarrow}). The strength is proportional to $(1 - J_0(\beta)) g_N$. Since $J_0(\beta) < 1$ for all $\beta$ there is no condition for which this magnetic field sensitivity vanishes \cite{sutherlandLaserfreeTrappedionEntangling2020}, meaning that the gate introduces susceptibility to magnetic field fluctuations.

For the four-tone scheme, we apply the transformation into the rotating frame of both the microwave drives. Assuming these two have the same Rabi frequency $\Omega_\mu$ and detuning $\delta$, we find
\be
H_N^{(4)} = \hbar g_N \cos(\omega_N t) J_0(\beta) \left( S_{z, \Uparrow} +  S_{z, \Downarrow}\right) \ .
\ee
and thus the qubit states remain insensitive to magnetic field noise, with the residual nuclear sensitivity remaining. This effectively relaxes the requirement for IDD in this scheme to be $J_0(\beta_{\Uparrow})=J_0(\beta_{\Downarrow})$, with $\beta_i=4\Omega_{\mu,i}/\delta_i$, which can be achieved by simply setting $\beta_{\Uparrow}=\beta_{\Downarrow}\equiv\beta$. It is notable here that the different values of $\phi$ used for the microwave tones on each of the spin flip transitions do not enter into the sensitivity to noise, while the choice of these values is critical to being able to perform the four-tone gate. 

\section{Parameter choices}

The 4-tone gate is inherently insensitive to magnetic fields if the modulation indices produced by the microwave drives on each transition are the same. In order to robustly ensure this condition, it seems favourable to work at $\beta= 3.05$, where $J_2(\beta) = 0.486$ is at a maximum, providing both the fastest possible gate time and insensitivity to fluctuations in $\Omega_\mu$. In the case that the microwave amplitudes can be well controlled and sufficient power is available, it might be advantageous to work at higher $\beta$, since the $S_y$ terms in equation \ref{eq:fullinteraction} provide the leading troublesome off-resonant effect. This is because $S_y$ does not commute with $S_z$ and therefore will cause leakage into the $\uparrow$ states. $\beta = 3.83$ corresponds to the zero of this unwanted term, which produces for the other terms $J_2(\beta) = J_0(\beta) = 0.402$. While the $J_0$ term is also undesired, it commutes with the desired interaction and thus can be counteracted effectively through appropriate dynamical decoupling control techniques. We note that in some species of ion it may be possible to detect the leakage, giving rise to improved error-correction properties \cite{sahayHighThresholdCodesNeutralAtom2023}. This is a topic which deserves further study in the current context.

\section{Conclusions}
We have discussed the implementation of nuclear spin qubits for quantum computing with trapped ions under typical operating conditions for Penning traps, in which the spin and nuclear degrees of freedom are largely separated by their different interaction strengths with the magnetic field, and only perturbatively coupled by the hyperfine interaction. We show that nuclear encodings in this regime suffer from slow bit-flip rates, but that all other operations can be performed at rates similar to those found in electron spin control without making compromises in magnetic field susceptibility using modifications of techniques demonstrated at low magnetic fields. In particular, we have shown that two-qubit phase gates, which are typically slow since they rely on coupling to field gradients, can be implemented at a speed similar to similar gates on electron spins by adapting the scheme of Sutherland et al. \cite{sutherlandLaserfreeTrappedionEntangling2020} using 4 microwave drives.

The 4-tone gate is distinct from previously proposed multi-qubit gates in trapped ions. While here described for nuclear spin qubits in Penning traps, the method is also feasible for applying phase gates to qubit transitions which occur at low field and are first-order magnetic field insensitive. This is advantageous since the magnetic-field independent qubits are also insensitive to low-frequency gradient fields \cite{ospelkausTrappedIonQuantumLogic2008}. This offers an alternative to concatenated dynamical decoupling methods which were applied previously \cite{nunnerichFastRobustLaserFree2025}. Although the four-tone gate shares some features with methods for applying state-resolved optical dipole forces to ions using quadrupole transitions \cite{baldwinHighfidelityLightshiftGate2021}, those methods are not easily transferred to the microwave domain, and would require gradient fields oscillating at much higher frequencies than those considered here. 

The high-field regime considered provides atomic systems for which all transitions are well resolved. The ground state structure of these atoms is such that for a given nuclear spin state, there is only one strongly allowed electron spin flip transition. This lack of spectator transitions provides an extremely clean environment for performing trapped-ion quantum computing which seems favorable for high fidelity operations, similar to the conditions found when using a ground state spin qubit in an even isotope, but with the advantage of the extended coherence time of the nuclear spin \cite{hughesTrappedionTwoqubitGates2025}. The large splitting of optical transitions (typically in the range of 10~GHz or more) provides the possibility to use optical pumping to operate on ions of the same species in a spectrally resolved manner, providing a means of sympathetic cooling while not having to deal with ions of different mass. Taken together with the demonstration of micro-fabricated Penning trap arrays \cite{jainPenningMicrotrapQuantum2024} and the recent developments in performing high fidelity operations \cite{hughesTrappedionTwoqubitGates2025} on increasingly large trapped ion systems \cite{ransfordHelios98qubitTrappedion2025}, this provides a promising avenue for scaling up quantum computation. 

\medskip \noindent \textbf{Acknowledgments}\\
We thank Daniel Kienzler for useful discussions, and for comments on the manuscript. We acknowledge support from the Swiss National Science Foundation (SNF) under SNSF Advanced Grant TMAG-$2$\_$225964$, and the Swiss Innovation Agency under Project No. 115.828 IP-ENG. M. M. was supported by the Rubicon project number 019.241EN.027 of the Dutch Research Council (NWO)

\medskip \noindent \textbf{Author contributions}\\
Considerations of nuclear qubits were made by M. M., J. F., S. J. and J. P. H. Two qubit gate mechanisms were originally proposed by J. P. H. and J. F., with follow up work from M. M. and J. A.. The paper was written by J. P. H., J. A., M. M. and J. F. with input from all authors.

\medskip\noindent\textbf{Competing interests}\\
J. P. H. is a scientific advisor of ZuriQ AG, a commercially oriented quantum computing company. J. P. H., J. F. and S. J. hold a financial interest in ZuriQ AG. J. P. H. also holds a financial interest in Qendra AG, and is a scientific advisor of that company. The other authors declare no competing interests. 

\bibliography{myrefs2.bib}

\end{document}